\documentclass[10pt,journal,compsoc]{IEEEtran}
\usepackage{array}
\usepackage{textcomp}
\usepackage{stfloats}
\usepackage{url}
\usepackage{verbatim}
\usepackage{cite}
\usepackage{balance}

\IEEEoverridecommandlockouts
\usepackage{amsmath}  
\usepackage[ruled,linesnumbered]{algorithm2e}
\usepackage{listings, xcolor}
\definecolor{verylightgray}{rgb}{.97,.97,.97}
\lstdefinelanguage{Solidity}{
  keywords=[1]{anonymous, assembly, assert, balance, break, call, callcode, case, catch, class, constant, continue, constructor, contract, debugger, default, delegatecall, delete, do, else, emit, event, experimental, export, external, false, finally, for, function, gas, if, implements, import, in, indexed, instanceof, interface, internal, is, length, library, log0, log1, log2, log3, log4, memory, modifier, new, payable, pragma, private, protected, public, pure, push, require, return, returns, revert, selfdestruct, send, solidity, storage, struct, suicide, super, switch, then, this, throw, transfer, true, try, typeof, using, value, view, while, with, addmod, ecrecover, keccak256, mulmod, ripemd160, sha256, sha3}, 
  keywordstyle=[1]\color{blue}\bfseries,
  keywords=[2]{address, bool, byte, bytes, bytes1, bytes2, bytes3, bytes4, bytes5, bytes6, bytes7, bytes8, bytes9, bytes10, bytes11, bytes12, bytes13, bytes14, bytes15, bytes16, bytes17, bytes18, bytes19, bytes20, bytes21, bytes22, bytes23, bytes24, bytes25, bytes26, bytes27, bytes28, bytes29, bytes30, bytes31, bytes32, enum, int, int8, int16, int24, int32, int40, int48, int56, int64, int72, int80, int88, int96, int104, int112, int120, int128, int136, int144, int152, int160, int168, int176, int184, int192, int200, int208, int216, int224, int232, int240, int248, int256, mapping, string, uint, uint8, uint16, uint24, uint32, uint40, uint48, uint56, uint64, uint72, uint80, uint88, uint96, uint104, uint112, uint120, uint128, uint136, uint144, uint152, uint160, uint168, uint176, uint184, uint192, uint200, uint208, uint216, uint224, uint232, uint240, uint248, uint256, var, void, ether, finney, szabo, wei, days, hours, minutes, seconds, weeks, years},  
  keywordstyle=[2]\color{teal}\bfseries,
  keywords=[3]{block, blockhash, coinbase, difficulty, gaslimit, number, timestamp, msg, data, gas, sender, sig, value, now, tx, gasprice, origin},  
  keywordstyle=[3]\color{violet}\bfseries,
  identifierstyle=\color{black},
  sensitive=false,
  comment=[l]{//},
  morecomment=[s]{/*}{*/},
  commentstyle=\color{gray}\ttfamily,
  stringstyle=\color{red}\ttfamily,
  morestring=[b]',
  morestring=[b]"
}
\usepackage{pifont}
\usepackage{diagbox}
\usepackage{subfigure}
\usepackage{oplotsymbl}
\usepackage{siunitx}
\usepackage{array,framed}
\usepackage{
   color,
   float,
   epsfig,
   wrapfig,
   graphics,
   graphicx,
 }
\usepackage{setspace}
\usepackage{amsfonts}
\usepackage{latexsym,fancyhdr,url}
\usepackage{enumerate}
\usepackage{algorithm2e}
\usepackage{algpseudocode}
\usepackage{graphics}
\usepackage{xparse} 
\usepackage{xspace}
\usepackage{multirow}
\usepackage{microtype}
\usepackage{threeparttable} 
\usepackage{fancyhdr}
\usepackage{hyperref}

\begin{document}
	
    \title{Unity is Strength: Enhancing Precision in Reentrancy Vulnerability Detection of Smart Contract Analysis Tools}
 
    \author{Zexu Wang,
        Jiachi Chen,
        Zibin Zheng,~\IEEEmembership{Fellow,~IEEE},
        Peilin Zheng,
        Yu Zhang,
        Weizhe Zhang
        \IEEEcompsocitemizethanks{\IEEEcompsocthanksitem Zexu Wang, Jiachi Chen, Zibin Zheng, Peilin Zheng are with School of Software Engineering, Sun Yat-sen University, China. \protect\\
                E-mail: \{wangzx97, zhengpl3\}@mail2.sysu.edu.cn
                
			E-mail: \{chenjch86, zhzibin\}@mail.sysu.edu.cn
        \IEEEcompsocthanksitem Weizhe Zhang, Yu Zhang are with School of Computer Science and Technology, Harbin Institute of Technology, China.\protect\\
			E-mail: \{yuzhang, wzzhang\}@hit.edu.cn 
        \IEEEcompsocthanksitem Zexu Wang, Yu Zhang and Weizhe Zhang are also affiliated with Peng Cheng Laboratory, China.
			
        \IEEEcompsocthanksitem Zibin Zheng is the corresponding author.}
	\thanks{Manuscript received     ; revised   
    }}
	

	\IEEEtitleabstractindextext{%
            \begin{abstract}
Reentrancy is one of the most notorious vulnerabilities in smart contracts, resulting in significant digital asset losses. However, many previous works indicate that current Reentrancy detection tools suffer from high false positive rates. Even worse, recent years have witnessed the emergence of new Reentrancy attack patterns fueled by intricate and diverse vulnerability exploit mechanisms. Unfortunately, current tools face a significant limitation in their capacity to adapt and detect these evolving Reentrancy patterns. Consequently, ensuring precise and highly extensible Reentrancy vulnerability detection remains critical challenges for existing tools. 

To address this issue, we propose a tool named ReEP, designed to reduce the false positives for Reentrancy vulnerability detection. Additionally, ReEP can integrate multiple tools, expanding its capacity for vulnerability detection. It evaluates results from existing tools to verify vulnerability likelihood and reduce false positives. ReEP also offers excellent extensibility, enabling the integration of different detection tools to enhance precision and cover different vulnerability attack patterns. We perform ReEP to eight existing state-of-the-art Reentrancy detection tools. The average precision of these eight tools increased from the original 0.5\% to 73\% without sacrificing recall. Furthermore, ReEP exhibits robust extensibility. By integrating multiple tools, the precision further improved to a maximum of 83.6\%. These results demonstrate that ReEP effectively unites the strengths of existing works, enhances the precision of Reentrancy vulnerability detection tools. 
\end{abstract}

        }
	
	\maketitle
	\IEEEdisplaynontitleabstractindextext


\section{Introduction}
\label{sec:intro}
\IEEEPARstart{S}{MART} contracts offer several unique features that distinguish them from traditional software programs, especially in finance and permission management scenarios~\cite{peckshield,slowmist}. Numerous smart contract vulnerabilities have been discovered through real-world attacks or theoretical analysis, including the Reentrancy vulnerability~\cite{ferreira2020aegis,ji2020deposafe}. Since the DAO attack in 2016, which resulted in the theft of approximately 150 million dollars in digital assets, the Reentrancy vulnerability has caused significant asset losses.

Reentrancy vulnerabilities arise when external (malicious) contracts exploit reentrant function characteristics to bypass permission control checks~\cite{zheng2023turn}. This allows external contracts to enter the same function multiple times, manipulate contract logic, and steal assets. 
A variety of technologies have been developed to detect Reentrancy vulnerabilities in smart contracts, which can be broadly divided into two categories, i.e., static analysis~\cite{21feist2019slither,16liao2022smartdagger,11ye2020clairvoyance} and dynamic analysis~\cite{10luu2016making,29mossberg2019manticore,15choi2021smartian,19so2021smartest}. Static analysis techniques often collect incomplete program state information, which can lead to false positives due to the loss of the state in contract interactions. Conversely, dynamic vulnerability detection models frequently struggle with deep-state search and comprehensive state analysis in cross-contract vulnerability scenarios. Zheng et al.~\cite{zheng2023turn} conducted a large-scale empirical study on existing popular Reentrancy vulnerability detection tools and found that these tools produced false positives as high as 99.8\%, with 55\% caused by \textit{incorrect permission control verification} and 41\% due to \textit{the lack of external contract function analysis}. Accurate permission control checks and cross-contract state analysis continue to challenge existing detection tools. Consequently, reducing the false positives of Reentrancy vulnerability detection remains a major topic in smart contract security research.

Despite significant research efforts directed toward detecting reentrancy vulnerabilities, the constant evolution of exploitation mechanisms has led to the emergence of new Reentrancy attack patterns in recent years. This complexity and variability necessitate a continuous expansion of vulnerability detection patterns within existing tools to ensure reliable detection of Reentrancy vulnerabilities. Unfortunately, many existing tools (such as Oyente~\cite{10luu2016making}, Osiris\cite{13torres2018osiris}, Manticore~\cite{29mossberg2019manticore}, etc.) have long struggled to effectively detect Reentrancy vulnerabilities in real-world contracts due to their outdated detection patterns.

To address the above challenges, we introduce ReEP, a tool designed to reduce false positives for Reentrancy vulnerability detection. ReEP evaluates results from existing tools and validates vulnerability likelihood to reduce false positives. When new vulnerability patterns emerge, ReEP integrates the corresponding detection tools to cover different vulnerability patterns. ReEP consists of two phases: \textit{target state search} and \textit{symbolic execution verification}. In the \textit{target state search} phase, ReEP uses program analysis to assist in pruning paths. 
ReEP performs program dependency analysis on the vulnerability functions provided by \textit{Origin Tools}, generating function sequences related to vulnerability triggering to guide the symbol execution. \textit{Origin Tools} refer to existing tools, such as Mythril~\cite{9mythril}, and Slither~\cite{Feist_Slither_Analyzer_2023}. Meanwhile, ReEP utilizes CFG Pruner to construct SMC-CFG (State Maximal Correlation CFG), which can optimize path traversal.
In the \textit{symbolic execution verification} phase, ReEP implements program instrumentation to collect and analyze path constraints, enabling cross-contract symbolic execution analysis. By utilizing symbolic execution to verify the reachability of these paths, we can reduce the false positives and enhance the precision of vulnerability detection for the \textit{Origin Tool}. Moreover, ReEP boosts strong extensibility by integrating new detection tools to expand the coverage of vulnerability detection patterns, thereby enhancing its capability to detect Reentrancy vulnerabilities.

We evaluated the effectiveness of ReEP by examing its ability to improve the detection precision of \textit{Origin Tools}, its capability to integrate multiple tools, and understanding the impact of each stage within the ReEP framework. The experimental results showed that when integrated with ReEP, the average precision of Origin Tools increased from 0.5\% to 73\%, significantly improving precision without sacrificing recall. 
Furthermore, ReEP is able to merge multiple Reentrancy detection tools to enhance its capabilities. By integrating six tools, ReEP achieves a peak precision of 83.6\%, while the best performance of the current state-of-the-art tools is only 31.8\%, demonstrating its effectiveness and extensibility in improving detection precision.
In addition, we conducted ablation experiments to understand how each stage within ReEP affects overall effectiveness. In general, ReEP provides a robust solution for improving the ability to detect Reentrancy vulnerabilities in smart contracts.

The main contributions of our work are as follows:
\begin{itemize}
    \item We designed a tool called ReEP to reduce the false positives for Reentrancy vulnerability detection. At the same time, it has strong extensibility in merging multiple tools to expand its capacity for vulnerability detection.
    \item We propose an approach that uses symbolic execution for verifying vulnerability path reachability. It combines program dependency analysis to guide path pruning, achieving efficient Reentrancy vulnerability verification.
    \item We applied ReEP to eight state-of-the-art Reentrancy vulnerability detection tools, experimental results show that it can significantly reduce the false positive rates and improve the precision of existing tools. 
    \item We publicize the ReEP’s source code and the experimental dataset at \href{https://github.com/ReEP-SC/ReEP}{https://github.com/ReEP-SC/ReEP}.
\end{itemize}

This paper is organized as follows. In Section~\ref{sec:background}, we describe some necessary background and explain the challenges faced in Reentrancy vulnerabilities through motivation examples. In Section~\ref{sec:methodology}, we introduce the workflow of our proposed method and delve into the technical details of ReEP. In Section~\ref{sec:evaluation}, we evaluate the performance of ReEP. We discuss threats to validity in Section~\ref{sec:discussion} and summarize the related work in Section~\ref{sec:related Work}. In Section~\ref{sec:conclusion}, we conclude the paper and outline future works.
\section{Background and motivation}
\label{sec:background}
In this section, we provide background knowledge and the motivation behind the design of the ReEP tool. Additionally, we summarize the challenges faced in Reentrancy vulnerability detection.

\subsection{Reentrancy}\label{sec:Reentrancy}
Since the DAO attack in 2016, detecting Reentrancy vulnerabilities has been a critical research topic in smart contract security. Attackers often exploit Reentrancy attacks to illegally acquire substantial amounts of digital assets, especially in DeFi applications where smart contracts manage significant volumes of digital assets. A Reentrancy attack is a type of malicious behavior that exploits a vulnerability in smart contracts, in which permission controls are inadequately checked when called by an external (malicious) contract. In such attacks, the attacker repeatedly enters the function through one function call to obtain considerable profit.
\begin{figure}[ht]
	\setlength{\abovecaptionskip}{0.cm}
	\begin{lstlisting}
  function withdraw(uint _amount) public {
    require(balance[msg.sender] >= _amount);
    (bool success, ) = msg.sender.call.value(_amount)("");
    require(success);
    balance[msg.sender] -= _amount;
  }
	\end{lstlisting}
	\caption{Simple example of Reentrancy}
	\label{SimpleExample1}
\end{figure}

Figure~\ref{SimpleExample1} shows a function named \textit{withdraw} that contains a Reentrancy vulnerability. In the \textit{withdraw} function, the contract first checks whether the \textit{caller} (represented by \textit{msg.sender}) has a sufficient balance (in L2). It then transfers the requested ether to the \textit{caller} (in L3) and deducts the transferred amount from the \textit{caller's} balance recorded in the user balance variable (in L5). However, Solidity introduces a special mechanism called the \textit{fallback function}, which can be used to execute code when the contract receives ether from other addresses. The \textit{fallback function} provides an opportunity for exploiting the Reentrancy vulnerability. In Figure~\ref{SimpleExample1}, the \textit{call.value()} (in L3) automatically invokes the \textit{fallback function} of the \textit{caller} contract, allowing the \textit{caller} to take control of the control flow. Attackers can deploy malicious code in the \textit{fallback function} to repeatedly call the \textit{withdraw()} function. Note that in the second invocation of \textit{withdraw()}, the code in L5 has not been executed since the invocation begins at the \textit{call.value()} in L3, and thus the user balance has not been changed at this time. As a result, the condition check (in L2) of the second invocation passes, and the victim contract will repeatedly transfer ether to the \textit{caller} until the contract's balance is drained.

\subsection{Motivation}\label{sec:Callcontextswitch}

We use the following example to illustrate practical applications and  key contributions of ReEP to smart contract development.

Alice, a smart contract developer, rigorously assesses her contracts for potential Reentrancy vulnerabilities before deploying them on Ethereum. To ensure a comprehensive analysis, she employs a variety of detection tools. However, different tools often report different vulnerability locations, compelling Alice to engage in thorough manual verification (as studies indicate that existing tools produce false positives as high as 99.8\%~\cite{zheng2023turn}). Furthermore, with the emergence of new Reentrancy attack patterns and corresponding tools, Alice needs to add them to her detection toolkit. However, the incorporation of these new tools may also generate new false positives, further increasing her workload.
\begin{figure}[htb]
	\setlength{\abovecaptionskip}{0.1cm}
	\centering
	\includegraphics[width=0.85\linewidth,height=32mm]{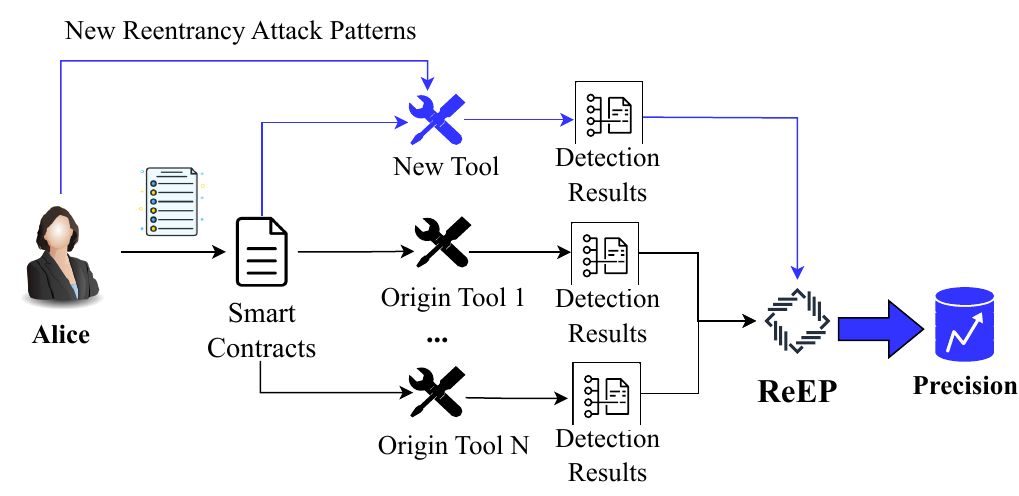}
	\caption{The use case of ReEP} 
	\label{fig:overview}
\end{figure}

At this stage, Alice can employ ReEP to efficiently reduce the false positives while consistently addressing the detection of new Reentrancy attack patterns, ultimately lightening her workload. ReEP is an automated verification tool, designed to validate the results of Reentrancy vulnerability detections from existing tools. It evaluates findings from multiple tools (\textit{Origin Tools}), verifying vulnerability likelihood to reduce false positives, thus enhancing precision. Moreover, ReEP has excellent extensibility. When new vulnerability patterns emerge, by incorporating the corresponding detection tools (\textit{New Tools}), Alice can further enhance Reentrancy vulnerability detection capabilities, ensuring broader coverage.

ReEP alleviates the manual verification of false positives, significantly cutting down Alice's workload. This underscores ReEP's practicality in streamlining audits and enhancing precision, suitable for large-scale smart contract detection tasks.

\subsection{Challenges}\label{sec:motivation2}
In this section, we investigate cases to uncover the main causes of false positives in current Reentrancy detection tools, and outline the related challenges.

\subsubsection{Lack of permission control check}
\vspace{-0.3cm} 
\begin{figure}[h]
	\setlength{\abovecaptionskip}{0.cm}
	\begin{lstlisting}
  modifier onlyOwner{
      require(msg.sender == owner);
      _;
  }
  ...
  function execute( address _to, uint _value, bytes _data) external onlyOwner {
      ...
      _to.call.value(_value)(data);
  }
	\end{lstlisting}
	\caption{Lack of permission control check}
	\label{MExample1}
\end{figure}
\vspace{-0.2cm} 

Figure~\ref{MExample1} shows a code snippet that leads to false positives in many detection tools. The main reason is that lacking of checks on the \textit{msg.sender's} permission control. Smart contract permission control mainly includes who has the right to call the function, what operations can be performed, and restrictions on contract operations. The form of permission control is diverse and extensive, mistakenly recognizing permission controls can readily lead to inaccurate detection results. Analyzing permission control correctly is the main challenge in improving tool detection accuracy.

\noindent{\bf Motivation:} Relying on matching the single pattern without analyzing permission control can easily result in false positives. For example, in Figure~\ref{MExample1}, the \textit{execute} function is modified with the \textit{onlyOwner} modifier, which restricts its execution to the contract owner. However, Mythril~\cite{9mythril} and Manticore~\cite{29mossberg2019manticore} did not capture this permission control logic and assumed that anyone could call this function, resulting in a false positive.

\noindent{\bf Challenge:} Different permission control mechanisms increase complexity and slow down path traversal. To improve detection efficiency and accuracy with symbolic execution, it's vital to optimize path selection and enable targeted analysis.

 \subsubsection{Unable to analyze execution logic of external contract functions}

Figure~\ref{MExample2} presents an instance from~\cite{zheng2023turn} where false positives occur due to the incapability of analyzing external contract function logic. The \textit{getTokenBal} function queries user balances through an external function \textit{balanceOf} (in L9), no state changes or transfers happen within \textit{balanceOf}. However, many tools can not fully analyse external functions, they often misidentify such contracts as having reentrancy vulnerabilities. 

\begin{figure}[htb]
	\setlength{\abovecaptionskip}{0.cm}
	\begin{lstlisting}
  contract ForeignToken {
      function balanceOf(address _owner) constant public returns (uint256);
      ...
  }
  contract Bitcash {
      ...
     function getTokenBal(address tokenAddr, address who) constant public returns (uint){
         ForeignToken t = ForeignToken(tokenAddr);
         bal = t.balanceOf(who);
         return bal;
     }
  }
	\end{lstlisting}
	\caption{Unable to analyze execution logic of external contract functions}
	\label{MExample2}
\end{figure}

\noindent{\bf Motivation:} Many detection tools struggle to understand how external contract functions work, especially in cross-contract interactions. For instance, in Figure~\ref{MExample2}, the \textit{balanceOf} function of the \textit{tokenAddr} address is called to check the balance of the \textit{who} address and update the \textit{bal} variable (in L9). However, these tools often can not determine if the external function involves transfers, so they rely on "state changes after external calls" to identify Reentrancy vulnerabilities, leading to false positives.

\noindent{\bf Challenge:} Cross-contract calls are challenging due to the difficulties in analyzing external contract functions. This often leads to incomplete program state analysis and false positives. 

\section{Methodology}
\label{sec:methodology}
In this section, we will introduce the workflow and delve into the technical details of ReEP.
\subsection{Overview}
\begin{figure}[ht]
	\setlength{\abovecaptionskip}{0.1cm}
	\centering
	\includegraphics[width=0.95\linewidth,height=43mm]{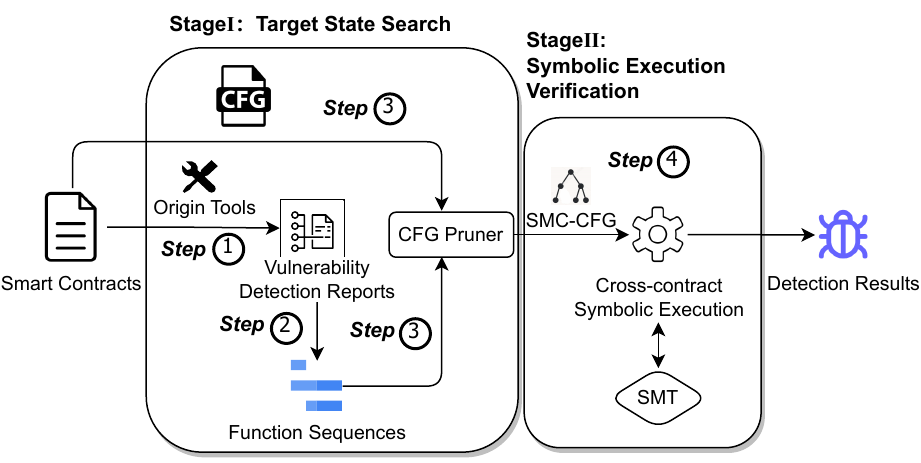}
	\caption{The workflow of ReEP} 
	\label{fig:workflow}
\end{figure}

We propose ReEP to verify vulnerability information in existing tools' detection reports, enhancing precision in Reentrancy vulnerability detection. As shown in Figure~\ref{fig:workflow}, the ReEP approach comprises two phases and four steps. It takes smart contract source code as input and produces detection results as output. In the first phase, \textit{Origin Tools} like Mythril~\cite{9mythril} are utilized to report vulnerability information, which includes the vulnerability's location and associated function. Program dependency analysis is then applied to generate the sequence of functions related to the vulnerability trigger, guiding the processes of symbolic execution. Following this, the CFG Pruner constructs the SMC-CFG (State Maximum Correlation Control Flow Graph) to optimize path traversal for symbolic execution. In the second phase, cross-contract interactions are monitored and analyzed to collect global path constraints and verify the reachability of vulnerability paths by accessing the constraint solver (SMT). This enables the determination of whether the vulnerability exists.

In general, \textit{Step\textcircled{1}}, \textit{Step\textcircled{2}}, \textit{Step\textcircled{3}} contribute to improve the efficiency of search, and \textit{Step\textcircled{4}} aims to improve the accuracy of detection.

\subsection{Stage \uppercase\expandafter{\romannumeral1}: Targeted State Search}

In order to improve the efficiency of state search, path pruning is essential for symbolic execution. ReEP achieves this by conducting program dependency analysis on vulnerable functions and generating function sequences to guide symbolic execution. Additionally, ReEP constructs the SMC-CFG to expedite path traversal during symbolic execution.

\subsubsection{Step\textcircled{1} Reentrancy Report Collection}\label{sec:reportcollection}
Initially, ReEP utilizes \textit{Origin Tools} to detect smart contracts and generate vulnerability detection reports, providing coarse-grained information on Reentrancy vulnerabilities. The information includes the location and function of the vulnerability, which ReEP further analyzes to generate the function sequence. The selection of \textit{Origin Tools} is based on the tool selection criteria outlined by Zheng et al.~\cite{zheng2023turn}, as these tools generate vulnerability detection reports containing the location and function information for Reentrancy vulnerabilities.

\subsubsection{Step\textcircled{2} Function Sequence Generation}\label{sec:vss}
To guide symbolic execution, ReEP analyzes functions and variables related to vulnerable functions through program dependency analysis to generate the sequence of functions. Therefore, it is necessary to collect information on the functions and variables related to the vulnerable functions in the contract. This information includes:
\begin{itemize}
    \item \textit{Target variables}: Variables of the functions from vulnerability detection reports, denoted as $V^{*}_{Target}$.
    \item \textit{Target-related variables}: Variables that have a program dependency relationship (including control dependencies or data dependencies) with $V^{*}_{Target}$, denoted as $V^{*}_{Target\_Related}$.
    \item \textit{Target functions}: Functions that read or assign variables that come from \textit{Target variables} or \textit{Target-related variables}, denoted as $F^{*}_{Target}$.
\end{itemize}

To collect the functions and variables related to the vulnerable function in the contract, ReEP performs a search based on whether they have program dependencies, including data and control dependencies, with the vulnerable functions. The following process is used:
\begin{itemize}
	\item When $V^{*}_{Target\_Related} =\emptyset$, search for the function that operates on the variables in $V^{*}_{Target}$, write the function into the set $F^{*}_{Target}$, and write the state variables of the function to the set $V^{*}_{Target\_Related}$.
	\item When $V^{*}_{Target\_Related} \neq\emptyset$, search for the function that operates on the variables in $V^{*}_{Target\_Related}$, write the function to the set $F^{*}_{Target}$, write the state variables of that function to the set $V^{*}_{Target\_Related}$, and continue repeating until $F^{*}_{Target}$ or $V^{*}_{Target\_Related}$ has no more new elements written to it, as $V^{*}_{Target} \cap  V^{*}_{Target\_Related} = V^{*}_{Target}$.
\end{itemize}

\begin{algorithm}[hb]  
  \caption{Generating Function Dependency Graph}\label{alg:generation}  
  \KwIn{$V^{*}_{Target\_Related}$ as V, $F^{*}_{Target}$ as F}
  \KwOut{FDG}
  \SetKwFunction{Main}{Main}
  \SetKwProg{Fn}{Function}{:}{}
  \Fn{\Main{V, F}}{
      FDG $\leftarrow $ empty graph\;
      \ForEach{$f_{i} $ in F}{
        \ForEach{$f_{j} $ in F}{
            \If{$modify(f_{i},V) \cap modify(f_{j},V) \neq \emptyset$}{
              $FDG.add\_edge(f_{i},f_{j},weight)$
            }
        }
      }
    \Return FDG\;  
  }
\end{algorithm}
To facilitate symbolic execution guidance, ReEP utilizes the FDG (Function Dependency Graph) to analyze the dependencies between vulnerable functions to generate the sequence of functions. The algorithm for generating the FDG is presented in Algorithm~\ref{alg:generation}, which takes $ V^{*}_{Target\_Related} $ and $ F^{*}_{Target} $ as inputs and produces the FDG as output.
The algorithm examines the dependency relationships between the contract's functions. Specifically, it determines whether there are same variables between two functions to generate the contract's FDG, as depicted in L3--L9 of Algorithm~\ref{alg:generation}. The edge weight is determined by the number of common variables shared by the two functions. By generating the FDG, ReEP can identify functions that are directly or indirectly related to the vulnerable functions, which enables the construction of a sequence of functions related to the execution of the vulnerable functions.

The sequence of functions is generated by sorting the functions in the FDG (Function Dependency Graph) according to their relevance. In the FDG, nodes represent functions, edges represent the existence of the same state variables, and the weight of edges represents the number of the common state variables. The weight of a node (function) is the sum of the weights of all its connecting edges, which indicate the frequency of operations on the state variables. Sorting nodes according to their weights generates the corresponding function sequence. 
To avoid uninitialized states, the nodes are sorted in ascending order of weight, creating the corresponding function sequence. 
Combining symbolic execution with function sequence guiding, ReEP achieves efficient access to functions related to the execution of the vulnerable functions, facilitating effective analysis of the critical path.

\subsubsection{Step\textcircled{3} Path Pruning}\label{sec:PathPruning}
To alleviate the problem of path explosion in symbolic execution, ReEP employs function sequences and the CFG Pruner to prune paths. Function sequences assist in eliminating irrelevant function access, while the CFG Pruner generates the SMC-CFG (State Maximum Correlation CFG) to prune the CFG of the function, thus enhancing the efficiency of symbolic execution. The SMC-CFG retains the CFG branch pointing to the key block where state updates occur and prunes irrelevant paths to minimize the cost of unnecessary path forking. At the CFG branch, it is checked whether the succeeding block is a key block. If all the succeeding blocks are key blocks, then all branches are preserved. Specifically, the CFG Pruner assigns weights to the jump edges of each basic block in the contract, reflecting the block's relevance to state updates. Blocks containing instructions that write or read state variables (\textit{SSTORE}, \textit{SLOAD}, and \textit{CALL}) are considered key blocks, while blocks without state operations or containing \textit{REVERT/INVALID} instructions are deemed irrelevant. If the condition for the vulnerability's existence is met, the symbolic execution path traversal is halted. By using the SMC-CFG, the path searching is accelerated by prioritizing the exploration of key blocks.

The algorithm for generating the SMC-CFG is presented in Algorithm~\ref{alg:generation_SMCCFG}. Initially, in L3--L9, all basic blocks in the CFG are traversed to calculate the weight of each edge, which is then recorded in the two-dimensional array SMC-CFG. In the \textit{Count\_weight} function in L12--L20, the opcode in each basic block is traversed. If the opcode is an instruction in \textit{Key\_instrs}, the weight of the corresponding connecting edge (JUMP\_W) for that basic block is incremented by 1. Simultaneously, the PC value is updated to the position of the first opcode in that block. By utilizing the \textit{Count\_weight} function, each basic block establishes weighted connections to its successor blocks, ultimately generating of the SMC-CFG.
\begin{algorithm}[h]  
  \caption{Generating the SMC-CFG }\label{alg:generation_SMCCFG}  
  \KwIn{the CFG of the contract}
  \KwOut{the SMC-CFG of the contract}

  \SetKwFunction{Main}{Main}
  \SetKwFunction{Count}{Count\_weight}
  \SetKwProg{Fn}{Function}{:}{}
  \Fn{\Main(CFG)}{
    SMC-CFG $\leftarrow$ $[][]$\;
    Blocks = getAllBlocks(CFG)\;
    \ForEach{block in Blocks}{
      \ForEach{blk in block.successors}{
        JUMP\_W, Fir\_pc = Count\_weight(blk)\;
        SMC-CFG[blk][Fir\_pc] = JUMP\_W\;
      }
    }
    \Return SMC-CFG 
  }
  \Fn{\Count(blk)}{
    JUMP\_W $\leftarrow$ 0,  PC $\leftarrow$ 0\;
    Key\_instrs = [SSTORE, SLOAD, CALL]\;
    \ForEach{opcode\ in\ blk.opcodes}{
      \If{opcode\ in\ Key\_instrs }{
        JUMP\_W++\;
        PC = blk.first\_opcode.pc\;
      }
    }
    \Return JUMP\_W, PC
  }
\end{algorithm}

\begin{figure}[h]
	\setlength{\abovecaptionskip}{0.1cm}
	\centering
	\includegraphics[width=0.70\linewidth,height=60mm]{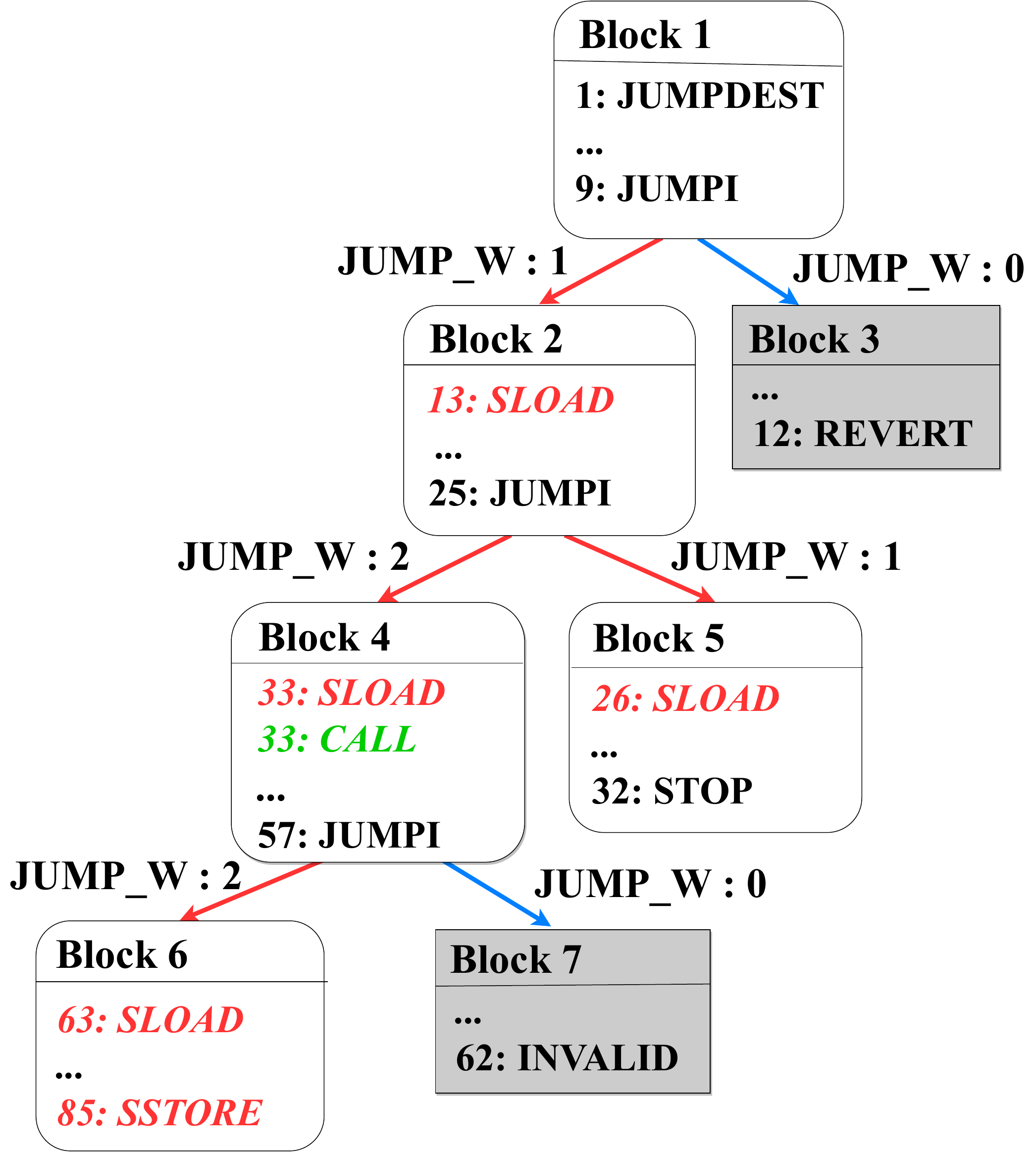}
	\caption{The SMC-CFG of the function \textit{withdraw}} 
	\label{fig:SMCCFG}
\end{figure}

Figure~\ref{fig:SMCCFG} illustrates a part of the SMC-CFG for the \textit{withdraw} function. The basic blocks in the SMC-CFG are represented by white rounded rectangles, while pruned blocks are shown as gray rectangles. The red solid lines indicate jump relationships between basic blocks, and the blue solid lines represent pruned jump relationships. The number on each edge denotes the weight value of the jump edge (\textit{JUMP\_W}). 
In Figure~\ref{fig:SMCCFG}, for Block 4, the \textit{JUMPI} opcode branch has jump edges with weight values of 0 and 2. During symbolic execution using the SMC-CFG, the jump edge with a weight value of 2 is explored, and the block with a weight value of 0 is omitted. If both succeeding blocks possess non-zero weight values, they are explored in the order of their weight values. By utilizing the jump edge weight values assigned to each block by the SMC-CFG, the path branching is prioritized by selecting the jump edge with a greater weight value, enabling symbolic execution to efficiently access paths with more state operations.

\subsection{Stage \uppercase\expandafter{\romannumeral2}: Symbolic Execution Verification}
To verify the existence of vulnerabilities, ReEP employs cross-contract symbolic execution to verify the reachability of the vulnerability path. It combines the SMC-CFG to accelerate path traversal. Additionally, it uses program instrumentation with function sequences to analyze the logic of functions and collect global path constraints.  Moreover, the SMT is accessed to validate the reachability of the path and determine the presence of any vulnerabilities.

\subsubsection{Step\textcircled{4} Cross-contract Symbolic Execution}\label{sec:CSE}
\begin{figure}[h]
	\setlength{\abovecaptionskip}{0.1cm}
	\centering
	\includegraphics[width=0.75\linewidth,height=60mm]{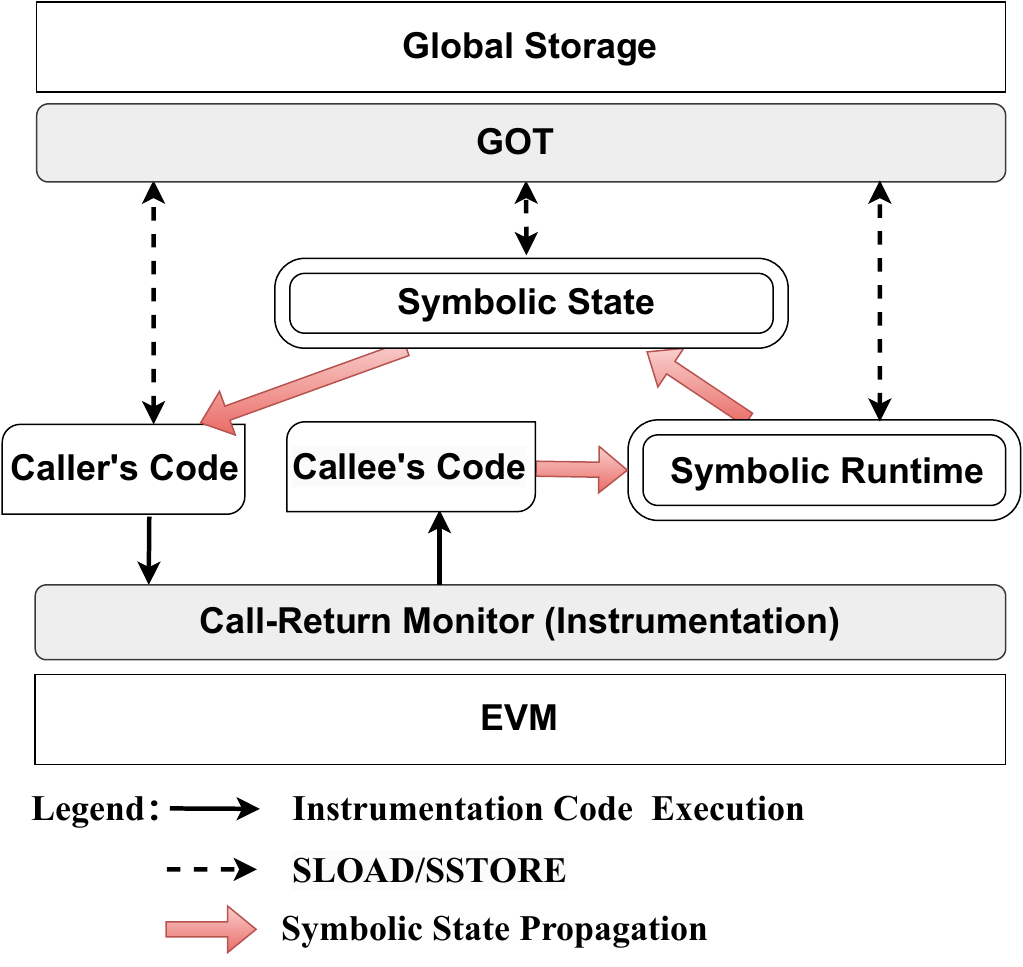}
	\caption{The overall process of cross-contract symbolic execution} 
	\label{fig:instrumentation}
\end{figure}

Analyzing the execution logic of functions and collecting global path constraints is of vital importance for cross-contract symbolic execution. Figure~\ref{fig:instrumentation} illustrates the overall process of cross-contract symbolic execution in ReEP. To identify and ensure the correct switching of different contract contexts (including \textit{msg} and \textit{storage}), ReEP employs the \textit{Call-Return Monitor}, a program instrumentation designed for cross-contract bytecode analysis, with function sequences to guide the switching of different contexts. The \textit{Global Storage} ensures the accurate writing and reading of distinct contract data, preventing data confusion arising from different contracts sharing a single storage. The \textit{Symbolic State Propagation} addresses the issue of symbolic parameters when calling across contracts. These modules work collectively to ensure the analysis of external functions and accurate collection of global path constraints.

\textbf{Call-Return Monitor.}
In order to ensure that the context of different contracts is switched correctly in cross-contract interaction, the instrumentation code needs to receive feedback on the start, and end positions, and return value information of the cross-contract call. As smart contracts use the Call-Return paradigm, which can be summarized into three patterns based on return values:

    \begin{figure}[H]
    \setlength{\abovecaptionskip}{0.cm}
    \begin{lstlisting}
    function Collect1(uint _am) public payable {
        if(balances[msg.sender]>=_am){
            if(msg.sender.call.value(_am)()){
                balances[msg.sender]-=_am;
                Log1.AddMessage(msg.sender,"Collect1");
    } } }
    \end{lstlisting}
    \caption{Call has no return value or does not use a return value}
    \label{MExample3-1}
    \end{figure}
    \textit{Call has no return value or does not use the return value}: the function called in the external contract (callee) does not return the value, or the return value is not used. This pattern is shown in L5 of the function \textit{Collect1} in Figure~\ref{MExample3-1}. The \textit{AddMessage} function in the \textit{Log1} contract does not return a value, and there is no need to pass the value across contracts.

    \begin{figure}[H]
    \setlength{\abovecaptionskip}{0.cm}
    \begin{lstlisting}
    function Collect2(uint _am) public payable{
        if(balances[msg.sender]>=_am){
            if (Log2.AddMessage(msg.sender,"Collect2")){
                if(msg.sender.call.value(_am)()){
                    balances[msg.sender]-=_am;		
    } } } }
    \end{lstlisting}
    \caption{Call uses the return value but does not assign it}
    \label{MExample3-2}
    \end{figure}
    \textit{Call uses the return value but does not assign it}: the function called in the external contract (callee) has the return value, which is used in the caller, but not assigned. This pattern is shown in L3 of the function \textit{Collect2} in Figure~\ref{MExample3-2}. The \textit{AddMessage} function in the \textit{Log2} contract is checked to see if it returns a value of True, but its return value is not assigned.

    \begin{figure}[h]
    \setlength{\abovecaptionskip}{0.cm}
    \begin{lstlisting}
    function Collect3(uint _am) public payable{
        if(balances[msg.sender]>=_am){
   success = Log3.AddMessage(msg.sender,"Collect3");
            if (success){
                if(msg.sender.call.value(_am)()){
                    balances[msg.sender]-=_am;
    } } } }
    \end{lstlisting}
    \caption{Call uses a return value and assigns it} 
    \label{MExample3-3}
    \end{figure}
    \textit{Call uses a return value and assigns it}: the function called in the external contract (callee) has the return value, which is used in the caller and assigned. 
    This pattern is shown in L3--L4 of the function \textit{Collect3} in Figure~\ref{MExample3-3}. The return value of the \textit{AddMessage} function in the \textit{Log3} contract is assigned to the \textit{success} variable.

The Call-Return Monitor employs static analysis of the bytecode stream to extract essential information about cross-contract function calls, including start and end positions, as well as the return values of external function calls. The Call-Return Monitor switches different contracts' \textit{storage} and \textit{msg} information based on the start and end positions and types of external calls. By analyzing the return value information of external function calls, it determines whether there is cross-contract parameter passing, ensuring the correct cross-contract interaction and avoiding state loss caused by cross-contract interactions.

\textbf{Global Storage.}
To ensure the accurate writing and reading of different contract data and to prevent data confusion, ReEP implements a global storage mechanism. Many dynamic detection tools fail to save data after the function call, leading to data loss issues in subsequent transaction execution and global state analysis, as the triggering of vulnerabilities frequently stems from multiple transactions. Moreover, different types of cross-contract calls, such as \textit{CALL}, \textit{DELEGATECALL}, \textit{CALLCODE}, and \textit{STATICCALL}, necessitate different contexts for the \textit{msg} and the \textit{storage}~\cite{42solidity}. Insufficient data or incorrect information can result in missed critical paths and inaccurate detection outcomes. To address this issue, ReEP combines the Global Storage with the GOT (Global Offset Table), which assists in locating and retrieving global data, thereby ensuring the correct storage and reading of various contract data.
\begin{figure}[htb]
	\setlength{\abovecaptionskip}{0.1cm}
	\centering
	\includegraphics[width=0.9\linewidth,height=35mm]{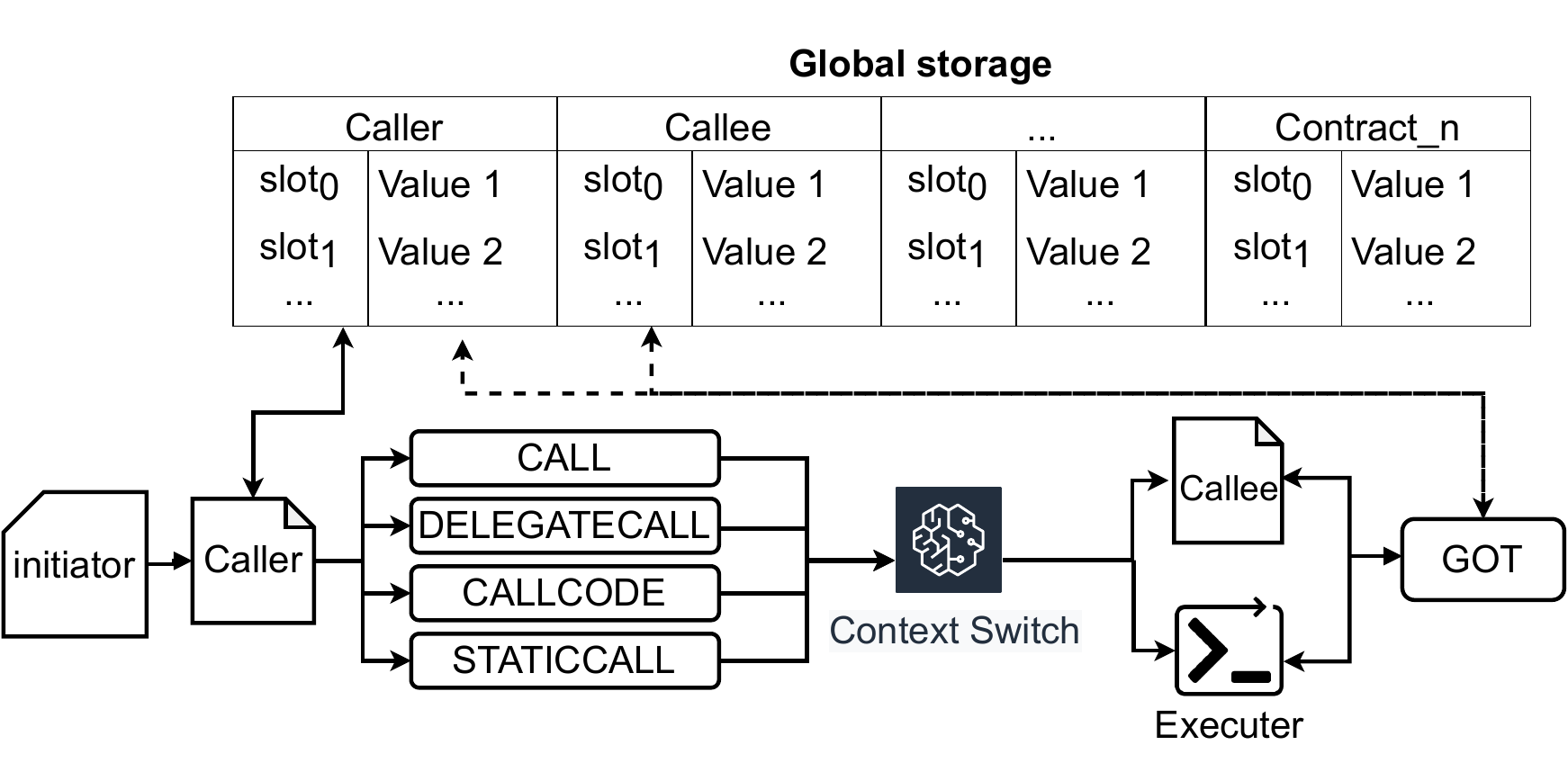}
	\caption{Global storage for call context switch} \label{fig:contextswitch}
\end{figure}

\begin{figure}[htbp]
	\setlength{\abovecaptionskip}{0.1cm}
	\centering
	\includegraphics[width=\linewidth,height=35mm]{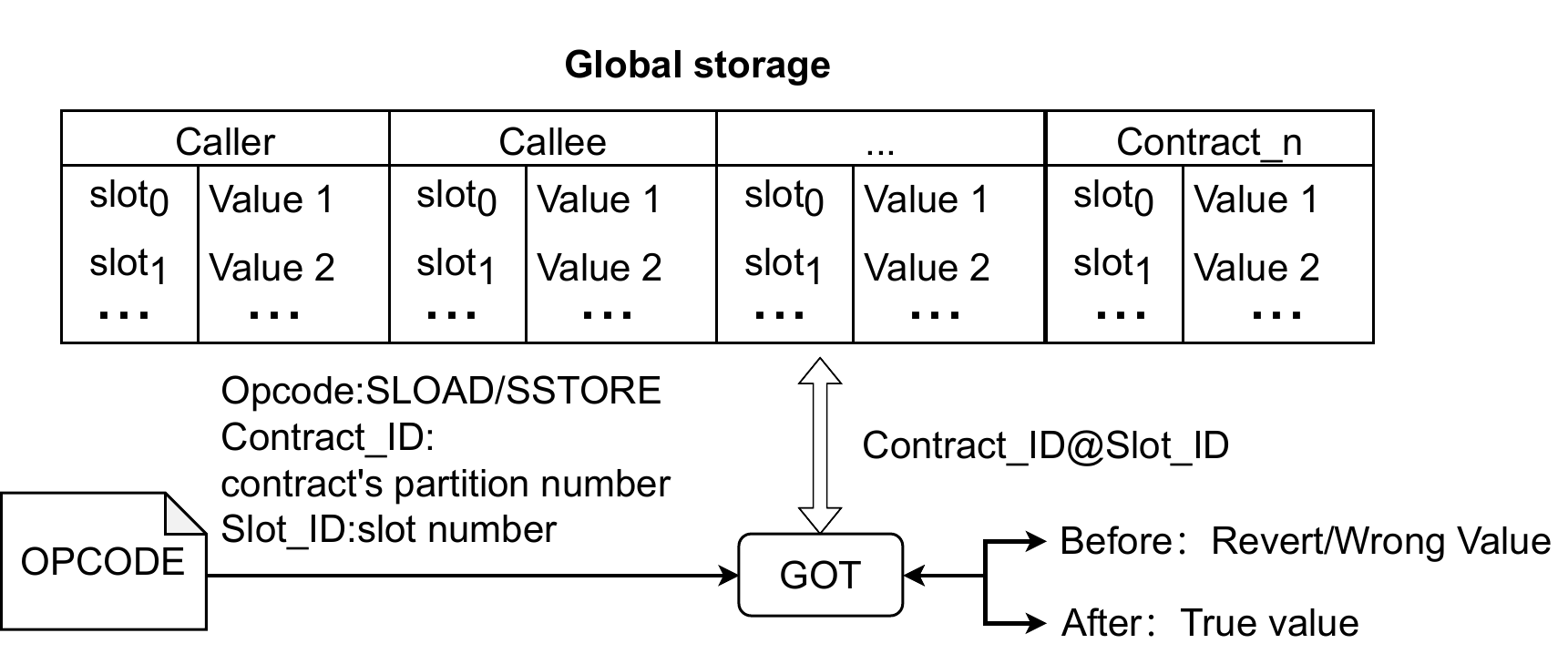}
	\caption{GOT (Global Offset Table)} \label{fig:GOT}
\end{figure}

Figure~\ref{fig:contextswitch} illustrates the functioning of global storage in the call context switch. 
The instrumentation code dynamically switches the contract context according to the various types of function calls between the caller and callee (including \textit{CALL}, \textit{DELEGATECALL}, \textit{CALLCODE}, and \textit{STATICCALL}) to ensure the accuracy of the cross-contract analysis.
Additionally, the GOT is used to determine the position of different contract data in the global storage. The design of the GOT is depicted in Figure~\ref{fig:GOT}, which aids in identifying the location of various contract data. The storage location of the global variable is calculated using the formula L = GOT($ Contract\_ID@Slot\_ID $), where $Contract\_ID$ represents the partition index of the contract in the global storage, and $Slot\_ID$ corresponds to the slot index in the storage of that contract. The instrumentation code monitors the \textit{SLOAD} and \textit{STORE} instructions and the corresponding stack top value in the bytecode, which are employed to compute $Slot\_ID$. By using GOT, true values can be obtained from global storage, avoiding issues such as reverts caused by incorrect or lost data.

\textbf{Symbolic State Propagation.}
To address the issue of passing symbolic parameters during cross-contract calls, ReEP returns the associated constraints of symbolic parameters as return values and passes them to the caller. During cross-contract transactions, the input parameters of the called function may be symbolic, and the output values or updated states may also be in the form of symbolic constraints, making immediate computation challenging. Symbolic state propagation transforms the contract operations on symbolic parameters into constraint expressions, which can be utilized for subsequent computations.

\begin{figure}[htb]
	\setlength{\abovecaptionskip}{0.1cm}
	\centering
	\includegraphics[width=0.88\linewidth,height=53mm]{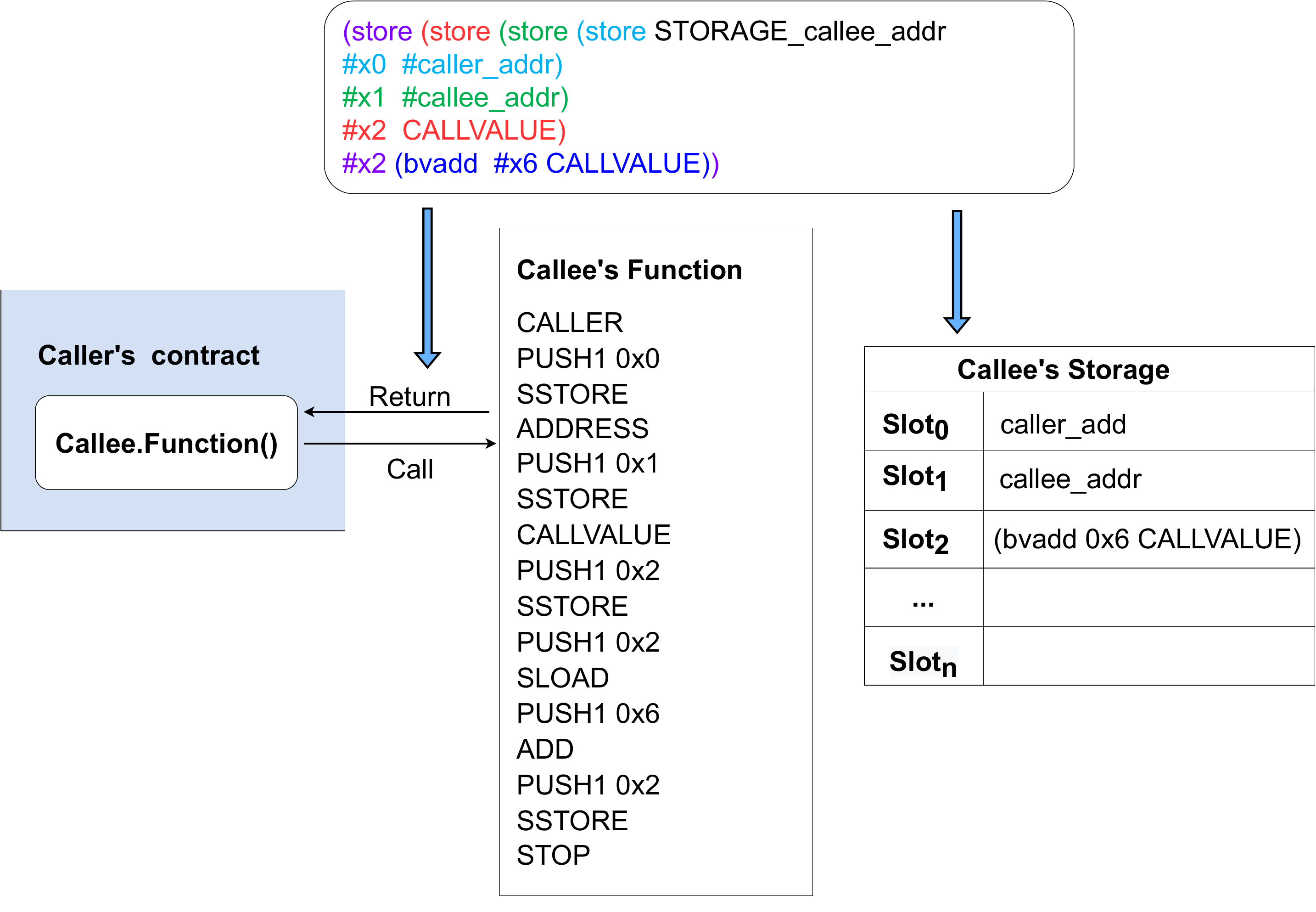}
	\caption{Storage and propagation of symbolic states} \label{fig:symbolic}
\end{figure}

ReEP converts the execution logic of external functions into symbolic constraint expressions and propagates them between different contracts. Z3's Bit-vector is employed to store symbolic constraint expressions, and the constraints of the input parameters (symbolic values) of the function are saved and propagated, resolving dependency issues of the program on symbolic parameters during symbolic execution. Figure~\ref{fig:symbolic} illustrates a simple example of a caller invoking a callee function with symbolic parameters, where the primary logic involves storing the sum of the transferred ether amount (\textit{CALLVALUE}) and 6. The bytecode block displayed in Figure~\ref{fig:symbolic} stores the \textit{caller\_addr} in $slot_{0}$, \textit{callee\_addr} in $slot_{1}$, and adds the transferred ether \textit{CALLVALUE} (a symbolic value) to 6, which is then stored in $slot_{2}$. ReEP converts the calculation operations on \textit{CALLVALUE} into constraint expressions for subsequent computation and storage. Since \textit{CALLVALUE} is a symbolic input parameter, the storage in $slot_{2}$ contains a symbolic constraint expression.

Path constraints are resolved by accessing the constraint solver (SMT). The completeness of the set of constraints during the path constraint search phase is essential for verifying the reachability of path constraints to determine the existence of vulnerabilities.
\section{Evaluation}
\label{sec:evaluation}
In this section, we evaluated the efficacy of ReEP by investigating its ability to enhance detection precision of the \textit{Origin Tools}, its ability to integrate multiple tools, and understand the impact of each stage within the ReEP framework. We address these considerations by answering the following research questions.
\begin{enumerate}[\bfseries RQ1.]
	\item How does ReEP improve the Reentrancy detection precision of the \textit{Origin Tools}?
 	\item What is the impact of ReEP on the recall rate?
        \item What is the extensibility of ReEP when merging multiple tools?
        \item What is the impact of different stages within ReEP?

\end{enumerate}

\subsection{Environment Setup and Dataset}
\label{sec:dataset}
\begin{table*}[htb]
	\setlength\tabcolsep{3pt}
	\setlength{\abovecaptionskip}{0.1cm}
	\begin{center}
		\caption{Statistics of detection results of the Origin Tool with ReEP}
		\label{tab:precision}
		\begin{threeparttable}
			\resizebox{0.98\textwidth}{!}{
\begin{tabular}{lllllllllllllllll}
\hline
\multicolumn{1}{c}{\textbf{Tool}}         & \multicolumn{2}{c}{\textbf{Oyente}}   & \multicolumn{2}{c}{\textbf{Mythril}}  & \multicolumn{2}{c}{\textbf{Securify1}} & \multicolumn{2}{c}{\textbf{Securify2}} & \multicolumn{2}{c}{\textbf{Smartian}} & \multicolumn{2}{c}{\textbf{Saifish}}  & \multicolumn{2}{c}{\textbf{Slither}}  & \multicolumn{2}{c}{\textbf{EThor}} \\
\multicolumn{1}{l|}{\textbf{}}            & Origin & \multicolumn{1}{l|}{Origin+} & Origin & \multicolumn{1}{l|}{Origin+} & Origin  & \multicolumn{1}{l|}{Origin+} & Origin  & \multicolumn{1}{l|}{Origin+} & Origin & \multicolumn{1}{l|}{Origin+} & Origin & \multicolumn{1}{l|}{Origin+} & Origin & \multicolumn{1}{l|}{Origin+} & Origin          & Origin+          \\ \hline
\multicolumn{1}{l|}{\textbf{\# TP}}       & 25     & \multicolumn{1}{l|}{25}      & 26     & \multicolumn{1}{l|}{26}      & 15      & \multicolumn{1}{l|}{22}      & 3       & \multicolumn{1}{l|}{3}       & 7      & \multicolumn{1}{l|}{7}       & 19     & \multicolumn{1}{l|}{21}      & 31     & \multicolumn{1}{l|}{31}      & 26              & 28               \\
\multicolumn{1}{l|}{\textbf{\# FP}}       & 488    & \multicolumn{1}{l|}{8}       & 15474  & \multicolumn{1}{l|}{8}       & 2372    & \multicolumn{1}{l|}{8}       & 2489    & \multicolumn{1}{l|}{5}       & 15     & \multicolumn{1}{l|}{5}       & 2270   & \multicolumn{1}{l|}{8}       & 4587   & \multicolumn{1}{l|}{9}       & 3269            & 9                \\
\multicolumn{1}{l|}{\textbf{\# Reported}} & 513    & \multicolumn{1}{l|}{33}      & 15500  & \multicolumn{1}{l|}{34}      & 2387    & \multicolumn{1}{l|}{30}      & 2492    & \multicolumn{1}{l|}{8}       & 22     & \multicolumn{1}{l|}{12}      & 2289   & \multicolumn{1}{l|}{29}      & 4618   & \multicolumn{1}{l|}{40}      & 3295            & 37               \\
\multicolumn{1}{l|}{\textbf{Precision}}   & 4.9\%  & \multicolumn{1}{l|}{75.8\%}  & 0.2\%  & \multicolumn{1}{l|}{76.5\%}  & 0.6\%   & \multicolumn{1}{l|}{73.3\%}  & 0.1\%   & \multicolumn{1}{l|}{37.5\%}  & 31.8\% & \multicolumn{1}{l|}{58.3\%}  & 0.8\%  & \multicolumn{1}{l|}{72.4\%}  & 0.7\%  & \multicolumn{1}{l|}{77.5\%}  & 0.8\%           & 75.7\%           \\ \hline
\end{tabular}
			}
		\end{threeparttable}
	\end{center}
\end{table*}

The experiments were conducted on a machine running Ubuntu 18.04.1 LTS and equipped with 16 cores (Intel(R) Xeon(R) Gold 5217). The experimental environment was set up by either downloading Docker images or manually building the tool with the help of construction manuals (for Securify2 and Smartian). To ensure that the experiments were not overly time-consuming, we followed the time budget settings from~\cite{zheng2023turn}, capping the maximum runtime to 120 seconds and keeping the parameters at their default settings. 

We adopted two datasets in our study to comprehensively evaluate ReEP's performance in detecting Reentrancy vulnerabilities. Dataset DB1 was employed to validate RQ1, RQ3, and RQ4, while dataset DB2 was used to validate RQ2.
\begin{enumerate}[\bfseries DB1.]
    \item \textit{Zheng et al.'s Dataset}~\cite{zheng2023turn}. Zheng et al. initially collected 230,548 verified contracts from Etherscan. These contracts were subsequently analyzed by using six state-of-the-art Reentrancy detection tools, and 21,212 contracts were flagged as potentially vulnerable to Reentrancy vulnerabilities by at least one of the tools. After that, two rounds of manual checks were conducted involving 50 participants, resulting in 34 contracts being confirmed as true positives (TP). 
    \item \textit{SmartBugs Dataset}~\cite{durieux2020empirical}. The widely used SmartBugs dataset contains 143 contracts. Among these contracts, 31 were identified as containing Reentrancy vulnerabilities. The inclusion of these labeled contracts allowed us to evaluate ReEP's recall performance in detecting Reentrancy vulnerabilities. 
\end{enumerate}

For tool selection, we adopted the criteria proposed by Zheng et al.~\cite{zheng2023turn}, selecting the same set of tools: Oyente~\cite{10luu2016making}, Mythril~\cite{9mythril}, Securify~\cite{23tsankov2018securify} (both V1 and V2 versions), Smartian~\cite{15choi2021smartian}, and Sailfish~\cite{bose2022sailfish}. Additionally, we included Slither~\cite{Feist_Slither_Analyzer_2023} and EThor~\cite{ethor} as part of the \textit{Origin Tools}, making a total of eight tools used to generate Reentrancy vulnerability detection reports. These selected tools are all presented at top software engineering or security conferences and cover various techniques, such as symbolic execution, formal verification, and fuzz testing. It is worth noting that some of these tools do not explicitly define Reentrancy vulnerabilities as ``Reentrancy”. For instance, Mythril reports two vulnerabilities related to Reentrancy, namely, \textit{external call to user-supplied address} and \textit{state access after external call}. To ensure consistency, we adopted the same classification criteria provided by Zheng et al.~\cite{zheng2023turn}, which offers a standardized classification scheme for these Reentrancy vulnerability detection tools.

\subsection{RQ1: Improvement on Precision}
We use DB1 to evaluate the impact of ReEP on improving the Reentrancy detection precision, we compared the results of eight \textit{Origin Tools} with and without ReEP. The results are provided in Table~\ref{tab:precision}, where \textit{TP} and \textit{FP} stand for True Positive and False Positive, respectively. \textit{Origin+} and \textit{Origin} represent the \textit{Origin Tool} with or without ReEP, and \textit{Precision} refers to the detection precision. The precision calculation formula is: $Precision(PRE) = TP/(TP+FP)$. 

As shown in Table~\ref{tab:precision}, a total of 21,212 Reentrancy cases  were reported from 230,548 contract detections by the \textit{Origin Tools}. At the same time, ReEP can significantly reduces false positives for all of them, especially for Mythril, where false positives dropped from 15,474 to 8. It is worth noting that DB1(\textit{Zheng et al.'s Dataset}) reported only 34 TP, while ReEP identified 7 additional contracts (a 20\% increase) with Reentrancy vulnerabilities that were missed in their manual checks. We notified the authors of the dataset, and they confirmed that these 7 contracts do contain Reentrancy vulnerabilities and subsequently updated their dataset. 

In conclusion, the results show that ReEP can significantly improve the precision of the \textit{Origin Tools}. Mythril exhibited the maximum increase in precision from 0.2\% to 76.5\%, while Smartian showed the minimum increase, also increasing from 31.8\% to 58.3\%. 
After using the ReEP, the average precision of all \textit{Origin Tools} rose by 72.5\%, from 0.5\% to 73\%, indicating that ReEP significantly improves the precision of the detection results for the \textit{Origin Tools}.

\textbf{Answer to RQ1:} The experimental results showed a significant improvement in the \textit{Origin Tools}'s precision when integrated with ReEP, escalating from 0.5\% to 73\% on average. The most significant improvement was observed in Mythril, with a remarkable increase of 76.3\%. These results highlight ReEP's exceptional capability in enhancing the precision of the \textit{Origin Tools}.

\subsection{RQ2: Impact on Recall}

\begin{table*}[t]
	\setlength\tabcolsep{3pt}
	\setlength{\abovecaptionskip}{0.1cm}
	\begin{center}
		\caption{Statistics of Detection Results on SmartBugs Dataset}
		\label{tab:RQ4}
		\begin{threeparttable}
			\resizebox{0.98\textwidth}{!}{
\begin{tabular}{lllllllllllllllll}
\hline
\multicolumn{1}{c}{\textbf{Tool}}       & \multicolumn{2}{c}{\textbf{Oyente}}    & \multicolumn{2}{c}{\textbf{Mythril}}   & \multicolumn{2}{c}{\textbf{Securify1}} & \multicolumn{2}{c}{\textbf{Securify2}} & \multicolumn{2}{c}{\textbf{Smartian}}  & \multicolumn{2}{c}{\textbf{Saifish}}   & \multicolumn{2}{c}{\textbf{Slither}}   & \multicolumn{2}{c}{\textbf{EThor}} \\
\multicolumn{1}{l|}{}                   & Origin  & \multicolumn{1}{l|}{Origin+} & Origin  & \multicolumn{1}{l|}{Origin+} & Origin  & \multicolumn{1}{l|}{Origin+} & Origin  & \multicolumn{1}{l|}{Origin+} & Origin  & \multicolumn{1}{l|}{Origin+} & Origin  & \multicolumn{1}{l|}{Origin+} & Origin  & \multicolumn{1}{l|}{Origin+} & Origin           & Origin+         \\ \hline
\multicolumn{1}{l|}{\textbf{\# TP}}     & 21      & \multicolumn{1}{l|}{21}      & 10      & \multicolumn{1}{l|}{10}      & 17      & \multicolumn{1}{l|}{17}      & 6       & \multicolumn{1}{l|}{6}       & 15      & \multicolumn{1}{l|}{15}      & 19      & \multicolumn{1}{l|}{19}      & 29      & \multicolumn{1}{l|}{29}      & 25               & 25              \\
\multicolumn{1}{l|}{\textbf{\# FP}}     & 43      & \multicolumn{1}{l|}{8}       & 48      & \multicolumn{1}{l|}{7}       & 31      & \multicolumn{1}{l|}{8}       & 47      & \multicolumn{1}{l|}{9}       & 19      & \multicolumn{1}{l|}{7}       & 24      & \multicolumn{1}{l|}{8}       & 38      & \multicolumn{1}{l|}{9}       & 55               & 9               \\
\multicolumn{1}{l|}{\textbf{\# FN}}     & 10      & \multicolumn{1}{l|}{10}      & 21      & \multicolumn{1}{l|}{21}      & 14      & \multicolumn{1}{l|}{14}      & 25      & \multicolumn{1}{l|}{25}      & 16      & \multicolumn{1}{l|}{16}      & 12      & \multicolumn{1}{l|}{12}      & 2       & \multicolumn{1}{l|}{2}       & 6                & 6               \\
\multicolumn{1}{l|}{\textbf{\# TN}}     & 69      & \multicolumn{1}{l|}{104}     & 64      & \multicolumn{1}{l|}{105}     & 81      & \multicolumn{1}{l|}{104}     & 65      & \multicolumn{1}{l|}{103}     & 93      & \multicolumn{1}{l|}{105}     & 88      & \multicolumn{1}{l|}{104}     & 74      & \multicolumn{1}{l|}{103}     & 57               & 103             \\ \hline
\multicolumn{1}{l|}{\textbf{Precision}} & 32.8\%  & \multicolumn{1}{l|}{72.4\%}  & 17.2\%  & \multicolumn{1}{l|}{58.8\%}  & 35.4\%  & \multicolumn{1}{l|}{68.0\%}  & 11.3\%  & \multicolumn{1}{l|}{40.0\%}  & 44.1\%  & \multicolumn{1}{l|}{68.2\%}  & 44.2\%  & \multicolumn{1}{l|}{70.4\%}  & 43.3\%  & \multicolumn{1}{l|}{76.3\%}  & 31.3\%           & 73.5\%          \\ \hline
\multicolumn{1}{l|}{\textbf{Recall}}    & 67.7\% & \multicolumn{1}{l|}{67.7\%} & 32.3\% & \multicolumn{1}{l|}{32.3\%} & 54.8\% & \multicolumn{1}{l|}{54.8\%} & 19.4\% & \multicolumn{1}{l|}{19.4\%} & 48.4\% & \multicolumn{1}{l|}{48.4\%} & 61.3\% & \multicolumn{1}{l|}{61.3\%} & 93.5\% & \multicolumn{1}{l|}{93.5\%} & 80.6\%          & 80.6\%         \\ \hline
\multicolumn{1}{l|}{\textbf{F1}}        & 44.2\% & \multicolumn{1}{l|}{70.0\%} & 22.5\% & \multicolumn{1}{l|}{41.7\%} & 43.0\% & \multicolumn{1}{l|}{60.7\%} & 14.3\% & \multicolumn{1}{l|}{26.1\%} & 46.2\% & \multicolumn{1}{l|}{56.6\%} & 51.4\% & \multicolumn{1}{l|}{65.5\%} & 59.2\% & \multicolumn{1}{l|}{84.1\%} & 45.0\%          & 76.9\%         \\ \hline
\end{tabular}
			}
		\end{threeparttable}
	\end{center}
\end{table*}

DB1 comprises 21,212 suspicious contracts selected from 230,548 verified contracts. It may not provide a fair evaluation of ReEP's recall impact, as these 21,212 contracts were marked as potentially susceptible to Reentrancy vulnerabilities by at least one of the Original tools. Therefore, we utilized DB2 (SmartBugs dataset), the most commonly used dataset, which consists of 31 contracts with Reentrancy vulnerabilities and 112 contracts without such vulnerabilities, to assess the recall impact. The recall calculation formula is: $Recall = TP / (TP+FN)$.

Table~\ref{tab:RQ4} presents the result statistics of ReEP on the SmartBugs dataset. As shown in the table, ReEP significantly enhances the precision and F1 scores of the \textit{Origin Tools}, while having no impact on Recall. Among the eight tools, Slither demonstrated the highest recall performance on SmartBugs, remaining unchanged at 93.5\%. This indicates that ReEP maintains the Recall of \textit{Origin Tools} while improving precision. 

Warning information is utilized for path pruning, combined with symbolic execution to verify path reachability, leading to fewer false positives and an enhanced F1 score, without affecting Recall of \textit{Origin Tools}.

\textbf{Answer to RQ2.}
ReEP utilizes warning information from \textit{Origin Tools} to guide path pruning and targeted analysis, including path symbol execution verification, resulting in a significant reduction of false positives in the detection results without compromising Recall rate.

\subsection{RQ3: Extensibility of ReEP}

The previous two RQs demonstrated that ReEP can improve precision by decreasing false positives (FP), but it cannot increase the recall rate. This is because ReEP relies on \textit{Origin Tools} to provide vulnerability information, and thus the ability of a single tool to report Reentrancy vulnerability might constrain ReEP's capability. Fortunately, this limitation can be countered by merging multiple tools, as a greater number of tools can provide more comprehensive vulnerability information.

In this RQ, we explore the detection efficiency of ReEP by combining multiple tools. The extensibility of ReEP was thoroughly assessed by fusing different sets of \textit{Origin Tools}, including combinations of two, four, six, and eight tools, resulting in a total of 127 unique combinations. All of the 127 combinations' experimental data are provided at our GitHub repository. When combining multiple tools, ReEP analyzes the detection outcomes of different tools, and determines the result by using a logical \textit{OR} operation. To clearly show the results, we categorized the results into three groups: \textit{Best\_combo}, \textit{Worst\_combo}, and \textit{Random\_combo}. Specifically, \textit{Best\_combo} represents the combination with the highest precision, \textit{Worst\_combo} corresponds to the combination with the lowest precision, and \textit{Random\_combo} includes randomly selected combinations.

 \vspace{-0.2cm}
\begin{figure}[htb]
	\setlength{\abovecaptionskip}{0.1cm}
	\centering
	\includegraphics[width=0.9\linewidth,height=33mm]{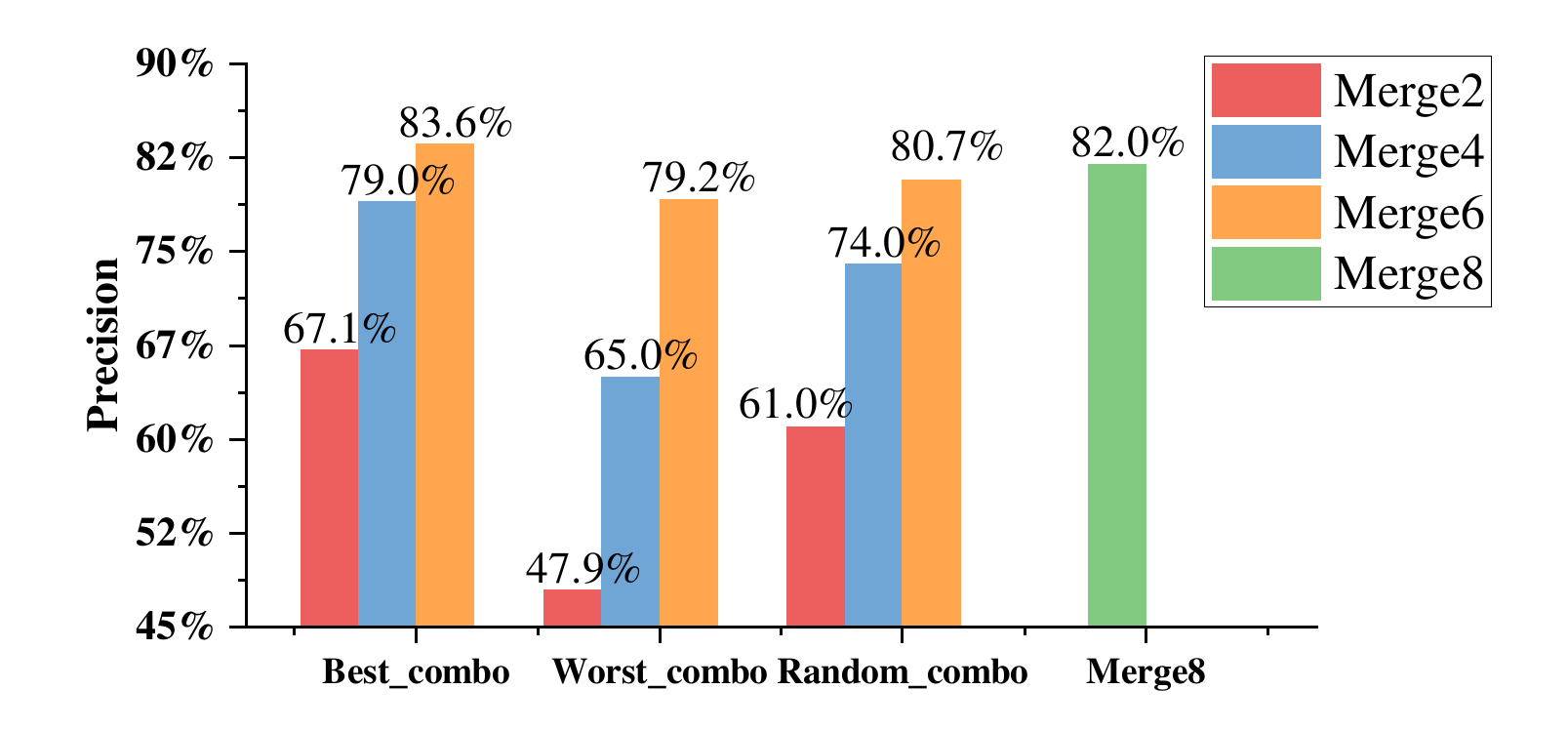}
	\caption{Comparison for merging multiple tools} \label{fig:comparsion-pre}       
\end{figure}
\vspace{-0.2cm}
Figure~\ref{fig:comparsion-pre} compares the precision of Merger2, Merge4, Merger6, and Merger8 with ReEP. Among the 28 combinations of Merger2, ReEP achieved the \textit{Best\_combo} by merging Mythril and Slither, resulting in the highest precision of 67.1\%. The \textit{Best\_combo} of Merger6 achieved the highest precision of 83.6\% among all the combinations, indicating the optimal performance achievable with these eight tools. This suggests that ReEP has already reached the best possible performance, and further increasing the number of merged tools does not result in an improvement in overall precision.
 Additionally, as the number of tools increases, the gap between the highest and lowest precision values decreases, resulting in more consistent results. Overall, ReEP effectively enhances detection precision by merging multiple tools, demonstrating its impressive extensibility.

\textbf{Answer to RQ3.}
Our experiments demonstrated that as the number of merged tools increased, the improvement in precision became more stable. The \textit{Best\_combo} of Merger6 achieved a peak precision of 83.6\%. Adding more tools did not significantly enhance the performance. These findings highlight the extensibility and effectiveness of ReEP in detecting Reentrancy vulnerabilities.

\subsection{RQ4: Impact of Different Stages in ReEP}

To evaluate the impact of each stage of ReEP on the overall effectiveness, we conducted comparative experiments on precision and time consumption for Merge6's \textit{Best\_combo} using three modes: \textit{Stage1} (Target State Search), \textit{Stage2} (Symbolic Execution Verification), and \textit{$Stage1\&2$}. In \textit{Stage1}, ReEP integrated the FDG and SMC-CFG analysis to prune the path, while \textit{Stage2} involved the verification of path accessibility through symbolic execution.

Table~\ref{tab:RQ3} presents a comparison of the detection results among the three modes, where \textit{Avg. Tool} represents the average detection results from the \textit{Origin Tools}. As evidenced by the table data, we can observe that \textit{Stage1} detected all the 41 Reentrancy vulnerabilities (TP) but have 556 false positives (FP). Stage 1 does not significantly improve the precision (from 0.5\% to 6.9\%) due to a high number of false positives. In contrast, \textit{Stage2's} symbolic execution verification remarkably reduces the number of false positives (FPs) to only 51, which is less than one-tenth of the FPs identified in \textit{Stage1}. Nonetheless, the number of true positives (TPs) detected by symbolic execution is 35, as symbolic execution may encounter path explosion, leading to the omission of certain TP contracts. The \textit{$Stage1\&2$} mode, which combines symbolic execution verification with path pruning, achieves the highest precision of 83.6\%.

\vspace{-0.1cm}
\begin{table}[h]
    \begin{center}
        \caption{Comparison of Detection Results for Different Modes}
        \label{tab:RQ3}
        \begin{threeparttable}
            \resizebox{\columnwidth}{!}{
            \tiny 
            \begin{tabular}{l|llll}
            \hline
            & \textbf{Avg. Tool} & \textbf{Stage1} & \textbf{Stage2} & \textbf{Stage1\&2} \\ \hline
            \textbf{\# TP} & 19 & 41 & 35 & 41 \\
            \textbf{\# FP} & 3870 & 556 & 51 & 8 \\ \hline
            \textbf{Precision} & 0.5\% & 6.9\% & 40.7\% & 83.6\% \\ \hline
            \textbf{Time (s)} & 8.18 & 1.3 & 23.8 & 15.6 \\ \hline
            \end{tabular}
            }
        \end{threeparttable}
    \end{center}
\end{table}
\vspace{-0.2cm}

Table~\ref{tab:RQ3} provides the average time consumption for the three modes. While \textit{Stage2} takes an average of 23.8 seconds, the processing time is reduced to 15.6 seconds in \textit{$Stage1\&2$}. The pruning in \textit{Stage1} plays a crucial role in enhancing ReEP's overall performance. Experimental observations indicate that \textit{Stage1} combines FDG with SMC-CFG pruning, leading to reduced running time and mitigated symbolic execution path explosion. Furthermore, it facilitates targeted path analysis, enabling deeper path searches and improving the overall precision of vulnerability detection.

\textbf{Answer to RQ4.}
The \textit{$Stage1\&2$} mode, by combining path pruning with symbolic execution verification, achieves efficient path reachability validation, resulting in the highest precision of 83.6\% when merging multiple tools.

\section{Threats to Validity}
\label{sec:discussion}
\subsection{Internal Validity}
The internal validity threats stem from the reliance on existing Reentrancy detection tools. While ReEP can improve precision by reducing false positives, it is not capable of detecting more Reentrancy cases than \textit{Origin Tools} (as observed in the experimental results of RQ2). Consequently, the capbilities of ReEP might be constrained by the limitations of \textit{Origin Tools}. Fortunately, a key strength of ReEP is its ability to merge multiple tools to enhance its detection capabilities. As demonstrated in RQ3, ReEP achieved a peak precision of 83.6\% by merging six tools, underlining its impressive extensibility. Even though new Reentrancy attack patterns emerge, ReEP can integrate new detection tools, effectively extending its capabilities to adapt to evolving vulnerabilities.

\vspace{-0.2cm}
\subsection{External Validity}
The external validity threats primarily stem from the inherent challenges associated with manual inspections of large datasets, which tend to be highly error-prone and time-consuming. The dataset \textit{DB1} comprises 230,548 verified contracts obtained from Etherscan, among which 21,212 contracts were flagged for Reentrancy vulnerabilities by state-of-the-art automated detection tools. Manually verifying this large-scale real-world dataset can be error-prone and time-consuming. To ensure the dataset's accuracy, the authors adopted a rigorous approach involving 50 participants, including graduate students and PhDs with extensive experience in smart contract research. These participants conducted two rounds of thorough checks on the detection results, minimizing bias and ensuring the reliability of the manual examination of the reentrant contracts detected by the tools. Utilizing tool evaluation to enhance the accuracy of the results. Additionally, by employing ReEP and integrating eight state-of-the-art tools, we successfully identified 7 additional real-world Reentrancy vulnerability contracts that were initially missed. Therefore, this dataset represents real-world smart contracts deployed on the Ethereum with Reentrancy vulnerabilities.


        \section{Related Work}
\label{sec:related Work}
\subsection{Symbolic Execution.}
Recent research on symbolic execution in smart contract security can be divided into two categories according to the effect: accuracy and efficiency.

To enhance the accuracy of symbolic execution in detecting security vulnerabilities, various tools such as Oyente~\cite{10luu2016making}, Mythril~\cite{9mythril}, Manticore~\cite{29mossberg2019manticore}, and Maian~\cite{31nikolic2018finding} conduct thorough analyses of contracts, and explore all possible paths to generate vulnerability reports. Sailfish~\cite{bose2022sailfish} utilizes a storage dependency graph (SDG) to detect vulnerabilities in contracts. SmartDagger~\cite{16liao2022smartdagger} constructs the cross-contract control flow graph from bytecodes, facilitating cross-contract vulnerability detection. However, blind path traversal can easily lead to imprecise results, especially in cross-contract analysis where different contracts sharing the same storage may cause data confusion. More importantly, many existing tools face a significant limitation in their capacity to adapt and detect evolving vulnerability patterns. ReEP distinguishes itself by validating existing tool results to reduce false positives and integrating different tools to detect various vulnerability patterns.

To improve the efficiency of symbolic execution for faster detection of security vulnerabilities, tools like Oyente~\cite{10luu2016making}, MPro~\cite{24zhang2019mpro}, and Mythril~\cite{9mythril} analyze smart contract bytecode. They apply corresponding path pruning strategies to expedite path traversal and mitigate path explosion issues. Smartian~\cite{15choi2021smartian}, SmartDagger~\cite{16liao2022smartdagger}, and Smartest~\cite{19so2021smartest} have optimized speed by refining search strategies for symbolic execution paths or employing heuristic algorithms. Additionally, Park~\cite{34zheng2022park} proposes a parallel symbolic execution-based approach to accelerate vulnerability detection. ReEP stands out by guiding symbolic execution based on the detection results of existing tools, achieving efficient path searching and traversal.

\vspace{-0.3cm}
\subsection{Dynamic and Static Detection.}
Traditional static detection techniques for smart contracts~\cite{21feist2019slither,16liao2022smartdagger,26schneidewind2020ethor,11ye2020clairvoyance,12tikhomirov2018smartcheck} have several limitations. The public nature of the blockchain, along with diverse permission control in smart contracts, complicates static vulnerability detection, often resulting in false positives. Additionally, unclear call relationships between functions require further exploration for effective target state identification triggering vulnerabilities.

Dynamic detection methods~\cite{10luu2016making,14ma2021pluto,29mossberg2019manticore,13torres2018osiris,24zhang2019mpro,15choi2021smartian,19so2021smartest,27su2022effectively,nguyen2020sfuzz}, relying on program testing and verification, enhance result reliability. However, the inability to access global state information and perform cross-contract analysis leads to path explosion and high resource consumption. Resource constraints and vulnerability identification limitations can further contribute to detection failures or errors.

FuzzSlice~\cite{fuzzslice} performs fuzz testing within a given time budget to eliminate potential false positives in static analysis. The difference is that ReEP validates suspicious vulnerability information from existing tools through symbolic execution, showcasing strong extensibilit and avoiding detection failures due to the inability to generate effective inputs. During the \textit{Target state search} phase, ReEP explores vulnerability target states to guide path pruning, eliminating irrelevant path branches to enhance path traversal efficiency. In the \textit{Symbolic execution verification} phase, combined with program instrumentation to achieve vulnerability state-guided symbolic execution verification. ReEP combines the strengths of both static and dynamic methods, which enables efficient identification and analysis of critical states to improve precision.

        \vspace{-0.2cm}
\section{Conclusion and Future Works}
\label{sec:conclusion}
In this paper, we present ReEP, a tool that effectively improves the precision of existing Reentrancy vulnerability detection tools. ReEP evaluates results from existing tools and validates vulnerability likelihood to reduce false positives. When new vulnerability patterns emerge, integrating the corresponding detection tools ensures reliable Reentrancy detection. 
By analyzing the program dependency relationships of vulnerability functions in the detection report, ReEP guides path pruning and symbolic execution. Cross-contract symbolic execution is employed to verify the reachability of vulnerability paths and confirm the existence of vulnerabilities.
We implemented and validated our tool with eight state-of-the-art detection tools. After applying ReEP, the average precision of these eight tools increased significantly from 0.5\% to 73\%. Furthermore, by merging six tools, the precision further improved, reaching a maximum of 83.6\%, while the best performance of the current state-of-the-art tools is only 31.8\%. These results demonstrate that ReEP effectively unites the strengths of existing works, enhances the precision of Reentrancy vulnerability detection tools, and efficiently identifies Reentrancy vulnerabilities in real-world scenarios.

In future work, we aim to expand the scope and capabilities of vulnerability detection by combining multiple technologies, covering a broader range of vulnerability types, and supporting the detection of bytecode, among other enhancements.

	\balance
        \bibliographystyle{IEEEtran}
        \nocite{*}
	\bibliography{refs}

\end{document}